\documentclass[]{mn2e}
\usepackage{times}

\title[PAHs and C/O ratios in planetaries]
{``PAH" emission bands in selected planetary nebulae:
a study of the behaviour with gas phase C/O ratio\footnote{based on observations with 
ISO, an ESA project with instruments funded by ESA member states (especially the 
PI countries: France, Germany, the Netherlands and the United Kingdom) and with the 
participation of ISAS and NASA}}

\author [M. Cohen \& M. J. Barlow]
{Martin Cohen\footnote{mcohen@astro.berkeley.edu}$^{1}$, M. J. Barlow$^{2}$\\
$^{1}$Radio Astronomy Laboratory, University of California,
    Berkeley, CA 94720\\
$^{2}$Department of Physics and Astronomy, University College London, 
  Gower Street, London WC1E 6BT, UK}

\date{ accepted . received ; in original form        }
\begin{document}

\maketitle

\begin{abstract}
Airborne and space-based low-resolution
spectroscopy in the 1980s discovered tantalizing quantitative 
relationships between the gas phase C/O abundance ratio in planetary nebulae 
(PNe) and the fractions of total far-infrared luminosity radiated by the 7.7 and 
11.3~$\mu$m bands (the C=C stretch and C-H bend, respectively), of polycyclic 
aromatic hydrocarbons.  Only a very small sample of nebulae was
studied in this context, limited by airborne observations of
the 7.7~$\mu$m band, or the existence of adequate $IRAS$ Low Resolution
Spectrometer data for the 11.3~$\mu$m band.  To investigate these trends further, we have
expanded the sample of planetaries available for this study using Infrared
Space Observatory (ISO) low-resolution spectra secured with the Short 
Wavelength Spectrometer (SWS) and the Long Wavelength Spectrometer (LWS).  
The new sample of 43 PNe, of which 17 are detected in PAH emission, addresses 
the range from C/O~=~0.2$-$13 with the objective of trying to delineate the pathways 
by which carbon dust grains might have formed in planetaries. 
For the 7.7-$\mu$m and 11.3-$\mu$m bands, we confirm that the ratio of
band strength to total infrared luminosity is correlated with the nebular
C/O ratio. Expressed in equivalent width terms, the cut-on C/O ratio
for the 7.7-$\mu$m band is found to be 0.6$^{0.2}_{0.4}$, in good accord with that 
found from sensitive ground-based measurements of the 3.3-$\mu$ band too.
\end{abstract}

\begin{keywords}
planetary nebulae, dust properties, abundances
\end{keywords}

\section{Introduction}
Emission features occur near 3.3, 6.2, 7.7, 8.7, 11.3~$\mu$m in infrared (IR)
nebular spectra (Russell, Soifer \& Merrill 1977; Russell, Soifer \& Willner
1977; Sellgren, Werner \& Dinerstein 1983; Aitken \& Roche 1982; Roche \&
Aitken 1986).  Airborne spectroscopy (Cohen et al. 1986,1989) shows these form a
generic spectrum, with the most intense peaks unobservable from the ground (6.2,7.7~$\mu$m);
all band intensities are correlated; and the 7.7-$\mu$m band tightly correlates with
gas phase C/O ratio in planetaries (i.e., for newly formed dust, in a circumstellar 
environment).  The features are attributed to vibrationally excited
polycyclic aromatic hydrocarbons (PAHs) and related materials (Duley \&
Williams 1981; Leger \& Puget 1984; Allamandola, Tielens \& Barker 1985;
Peeters et al. 2002).  The PAHs are regarded as permeating almost every phase of 
the interstellar medium (ISM)(Allamandola, Hudgins, \& Sandford 1999).
Various components are recognized (Allamandola, Tielens \& Barker 1989; Tielens
1993): the narrow features are carried by molecular-sized PAHs (50 C-atoms);
larger PAH clusters (500 C-atoms) yield the plateaus underlying the narrow
features; while the 25 and 60-$\mu$m ``cirrus" emission originates from
amorphous carbon grains, perhaps built
from PAHs, of size 5000 to 50,000 C-atoms.  This hypothesis is supported by the
obvious link between carbon abundance and the bands (Barlow 1983; Cohen et al.
1986,1989; Casassus et al. 2001), 
and the tight correlation between the 6.2 and 7.7-$\mu$m bands (both from
C=C skeletal modes).  The 11.3-$\mu$m feature is due to out-of-plane bending
vibrations of peripheral H atoms attached to aromatic molecular units; its
precise frequency depends on the number of adjacent H atoms on each edge ring
(Bellamy 1958).  The observed 11.3 and 12.7-$\mu$m bands and the 10.5-14~$\mu$m plateau of
emission (Cohen, Tielens \& Allamandola 1985) are characteristic of aromatic
rings with nonadjacent, or 2 or 3 adjacent, peripheral H atoms (see also
van Diedenhoven et al. 2004).  With the advent of ISO (Kessler et al. 1996) 
spectra, additional features attributed to PAHs have been
identified, such as the 16.4-$\mu$m peak and plateau, attributed by Moutou et al.
(2000) to C-triple bonds. 

Using exclusively the $IRAS$ Low Resolution Spectrometer (LRS) data base,
Volk \& Cohen (1990) sought a correlation of the 11.3-$\mu$m band with C/O, finding
an apparently constant value of 11.3-$\mu$m strength for C/O$>$1, with an abrupt
transition near C/O~=~1, the boundary between O- and C-rich nebulae.  Any
relationship between the 7.7-$\mu$m or 11.3-$\mu$m bands and gas phase C/O may also hold clues to
the mechanisms of dust formation in planetaries, and might even elucidate the
role of peripheral hydrogen atoms.

Understanding carbon dust formation has wide significance in astronomy, because
AGB stars, the precursors of C-rich PNe, are the dominant formation sites known for 
refractory grains 

\onecolumn
\begin{table*}
\begin{center}
\caption{ISO SWS and LWS data of PNe gathered for this study}
\begin{tabular}{cllcrcrl}
\hline
Object& TDT& AOT& Date& Integration(s)& LWS off?& Diam.($^{\prime\prime}$)& Diameter reference\\ 
\hline
BD+30$^\circ$3639& 35501412& LWS01& 06-Nov-1996& 1266& N&   8.0&       Acker et al. (1992)\\
BD+30$^\circ$3639& 35501531& SWS01& 06-Nov-1996& 1140& ...     &			  \\
BD+30$^\circ$3639& 86500540& SWS01& 29-Mar-1998& 3454& ...     &			 \\
CPD-56$^\circ$8032& 08401538& LWS01& 09-Feb-1996& 1552& N&  1.9&       De Marco et al. (1997)       \\
CPD-56$^\circ$8032& 13602083& SWS01& 01-Apr-1996& 3462& ...&   &			   	     \\
CPD-56$^\circ$8032& 27301339& SWS01& 16-Aug-1996& 1140& ...&   &			  	     \\
Cn\,1-5& 47101650& SWS01& 01-Mar-1997& 1912& N&	    27&        Schwarz et al. (1992)\\
Cn\,1-5& 47101651& LWS01& 01-Mar-1997& 1052& Y&	       &			   \\
Cn\,1-5& 48500404& LWS01& 15-Mar-1997& 1908& Y&	       &			   \\
Hb\,12& 43700330& SWS01& 26-Jan-1997& 1912& ...&            0.4&       Zhang \& Kwok (1993)         \\
Hb\,12& 57101028& LWS01& 09-Jun-1997& 1318& Y&                 &			   	     \\
He\,2-113& 07903229& LWS01& 04-Feb-1996& 1554& Y&           1.3&       De Marco et al. (1997)       \\
He\,2-113& 07903307& SWS01& 04-Feb-1996& 1044& ...           &			   	     \\
He\,2-113& 43400768& SWS01& 23-Jan-1997& 1912& ...&            &			   	     \\
He\,2-113& 60701891& LWS01& 15-Jul-1997& 620& Y&               &			   	     \\
He\,2-113& 66900121& LWS01& 14-Sep-1997& 2228& Y&              &			   	     \\
He\,2-131& 07902010& SWS01& 04-Feb-1996& 1044& ...&         6.0&       Zhang \& Kwok (1993)         \\
He\,2-131& 27301830& LWS01& 16-Aug-1996& 1268& N&              &			   	     \\
Hu\,1-2& 35801255& SWS01& 09-Nov-1996& 1912& ...&           8.2&       Zhang \& Kwok (1993)         \\
Hu\,1-2& 35801256& LWS01& 09-Nov-1996& 710& Y&              	      & 			    \\
Hu\,2-1& 13400705& SWS01& 30-Mar-1996& 1834& N&             1.8&       Acker et al. (1992)  	     \\
IC\,3568& 21304921& LWS01& 17-Jun-1996& 1256& N&            12 &       Zhang \& Kwok (1993) 	     \\
IC\,3568& 21304923& SWS01& 17-Jun-1996& 1140& ...&          	      & 			    \\
IC\,418& 68900805& LWS01& 04-Oct-1997& 3430& Y&             12 &       Acker et al. (1992)          \\
IC\,418& 82901301& SWS01& 22-Feb-1998& 1912& ...&           	      & 			    \\
IC\,418& 86801205& LWS01& 01-Apr-1998& 3428& Y&             	      & 			    \\
IC\,4406& 43600728& SWS01& 25-Jan-1997& 1912& ...&          35 &       Zhang \& Kwok (1993) 	     \\
IC\,4406& 44700327& LWS01& 05-Feb-1997& 1318& N&            	      & 			    \\
IC\,4997& 31901334& SWS01& 30-Sep-1996& 1140& ...&          1.6&       Zhang \& Kwok (1993)         \\
IC\,4997& 37400215& LWS01& 24-Nov-1996& 1858& Y&            	      & 			    \\
IC\,5117& 36701822& LWS01& 18-Nov-1996& 1858& N&            1.4&       Zhang \& Kwok (1993)         \\
IC\,5117& 36701824& SWS01& 18-Nov-1996& 1140& ...&          	      & 			    \\
M\,1-42& 48500234& LWS01&  15-Mar-1997& 1316& N&            9.0&       Zhang \& Kwok (1993) 	     \\
M\,1-42& 48500235& LWS01&  15-Mar-1997& 1318& N&            	      & 			    \\
M\,1-42& 70302306& SWS01&  19-Oct-1997& 1912& ...&          	      & 			    \\
M\,2-36& 70302403& SWS01&   19-Oct-1997& 1912& ...&         7.0&       Zhang \& Kwok (1993) 	     \\
M\,4-18& 83801755& SWS01& 02-Mar-1998& 1912& ...&           3.7&       De Marco \& Crowther (1999)  \\
Mz\,3& 08402133& LWS01& 09-Feb-1996& 1552& Y&               25 &       Zhang \& Kwok (1993)        \\
Mz\,3& 27300834& SWS01& 15-Aug-1996& 1140& ...&             	      & 			    \\
NGC\,2346& 71602536& LWS01&  01-Nov-1997& 2230& N&          55 &       Zhang \& Kwok (1993)         \\
NGC\,2346& 71602537& SWS01&  01-Nov-1997& 1912& ...&        	      & 			    \\
NGC\,2440& 72501762& SWS01& 10-Nov-1997& 1912& ...&         18 &       Zhang \& Kwok (1993)         \\
NGC\,3918& 26700720& LWS01& 09-Aug-1996& 1268& Y&           16 &       Zhang \& Kwok (1993)         \\
NGC\,3918& 29900201& SWS01& 10-Sep-1996& 1140& N&           	      & 			    \\
NGC\,40& 30003803& SWS01& 12-Sep-1996& 3454& ...&           36.4&      Zhang \& Kwok (1993)	     \\
NGC\,40& 44401917& SWS01& 02-Feb-1997& 1912& ...&              &			   	     \\
NGC\,40& 47300616& LWS01& 03-Mar-1997& 1318& Y&                &			   	     \\
NGC\,5189& 31800124& LWS01&      29-Sep-1996& 1266& Y&      140&       Zhang \& Kwok (1993) 	     \\
NGC\,5189& 31800125& SWS01&      29-Sep-1996& 1140& ...&    	      & 			   \\
NGC\,5315& 07902104& SWS01& 04-Feb-1996& 1044& ...&         6.1&       Zhang \& Kwok (1993)	     \\
NGC\,5315& 28001926& LWS01& 23-Aug-1996& 1268& Y&              &			   	     \\
NGC\,5315& 43600267& SWS01& 25-Jan-1997& 1912& ...&            &			   	     \\
NGC\,6153& 08402635& LWS01& 09-Feb-1996& 1552& Y&           23 &       Zhang \& Kwok (1993)         \\
NGC\,6153& 08402713& SWS01& 09-Feb-1996& 1044& ...&         	      & 			    \\
NGC\,6153& 45901470& SWS01& 17-Feb-1997& 1912& ...&         	      & 			    \\
NGC\,6210& 30400331& SWS01& 15-Sep-1996& 1912& ...&         16 &       Zhang \& Kwok (1993)         \\
NGC\,6210& 30400332& LWS01& 15-Sep-1996& 650& Y&            	      & 			    \\
NGC\,6210& 46300706& LWS01& 21-Feb-1997& 586& Y&            	      & 			    \\
NGC\,6302& 09400716& SWS01& 19-Feb-1996& 6528& ...&         45 &       Zhang \& Kwok (1993)         \\
NGC\,6302& 28901940& LWS01& 01-Sep-1996& 1266& Y&           	      & 			    \\
NGC\,6369& 31100910& LWS01& 23-Sep-1996& 1340& N&           29&        Zhang \& Kwok (1993)         \\
NGC\,6369& 45601901& SWS01& 14-Feb-1997& 1140& ...&            &                                  \\
\hline
\end{tabular}
\end{center}
\end{table*}

\begin{table*}
\begin{center}
\contcaption{}
\begin{tabular}{cllcrcrl}
\hline
Object& TDT& AOT& Date& Integration(s)& LWS off?& Diameter($^{\prime\prime})$& Diameter reference\\
\hline
NGC\,6445& 48700507& SWS01& 17-Mar-1997& 1912& ...&      33 &	    Zhang \& Kwok (1993)  	 \\
NGC\,6445& 48700508& LWS01& 17-Mar-1997& 640& Y&         	   &			        \\
NGC\,6537& 47000722& SWS01& 28-Feb-1997& 1912& ...&      75 &	    Schwarz et al. (1992)	 \\
NGC\,6537& 47000802& LWS01& 28-Feb-1997& 1318& Y&        	   &			        \\
NGC\,6537& 70300475& SWS01& 18-Oct-1997& 3454& ...&      	   &			        \\
NGC\,6543& 02400714& SWS01& 11-Dec-1995& 6544& ...&      20 &	    Zhang \& Kwok (1993)  	 \\
NGC\,6543& 02400807& SWS01&      11-Dec-1995& 3484& ...& 	   &			        \\
NGC\,6543& 02400910& SWS01&      11-Dec-1995& 1096& ...& 	   &			        \\
NGC\,6543& 02800908& SWS01&      15-Dec-1995& 1094& ...& 	   &			        \\
NGC\,6543& 03201202& SWS01&      19-Dec-1995& 1860& ...& 	   &			        \\
NGC\,6543& 25500701& LWS01&      28-Jul-1996& 1266& ...& 	   &			        \\
NGC\,6572& 30901603& LWS01&      21-Sep-1996& 1266& Y&   14.&	    Zhang \& Kwok (1993) 	 \\
NGC\,6572& 31901125& SWS01&     30-Sep-1996& 1140& ...&  	   &			        \\
NGC\,6720& 17601005& LWS01&  11-May-1996& 4321& N&       70 &	    Zhang \& Kwok (1993)  	 \\
NGC\,6720& 36600206& LWS01&      16-Nov-1996& 1268& N&   	   &			        \\
NGC\,6720& 36600207& SWS01&      16-Nov-1996& 1140& ...& 	   &			        \\
NGC\,6741& 13401806& SWS01&  30-Mar-1996& 1062& ...&     7.8&	    Zhang \& Kwok (1993) 	 \\
NGC\,6790& 13401107& SWS01& 30-Mar-1996& 1062& ...&      1.8&	    Zhang \& Kwok (1993) 	 \\
NGC\,6790& 13401608& LWS01& 30-Mar-1996& 1330& N&        	   &			        \\
NGC\,6826& 30201113& LWS01& 14-Sep-1996& 1266& N&        25 &	    Zhang \& Kwok (1993)  	 \\
NGC\,6826& 30201114& SWS01& 14-Sep-1996& 1140& ...&      	   &			        \\
NGC\,6884& 13901709& SWS01& 04-Apr-1996& 1834& ...&      6.0&	    Zhang \& Kwok (1993) 	 \\
NGC\,6886& 13400810& SWS01& 30-Mar-1996& 1062& ...&      7.4&	    Zhang \& Kwok (1993) 	 \\
NGC\,6891& 37600943& SWS01& 27-Nov-1996& 1912& ...&     12.6&      Zhang \& Kwok (1993)	 \\
NGC\,6891& 37600944& LWS01& 27-Nov-1996& 1166& Y&        	   &			        \\
NGC\,7009& 34400517& LWS01& 25-Oct-1996& 1268& N&        27 &	    Zhang \& Kwok (1993)   	 \\
NGC\,7009& 34400518& SWS01& 25-Oct-1996& 1140& ...&      	   &			        \\
NGC\,7009& 73801242& SWS01& 23-Nov-1997& 1912& ...&      	   &			        \\
NGC\,7027& 02401183& SWS01& 11-Dec-1995& 6542& ...&      15 &	    Acker et al. (1992) 	 \\
NGC\,7027& 23001256& SWS01& 04-Jul-1996& 1140& ...&      	   &			        \\
NGC\,7027& 23001357& SWS01& 04-Jul-1996& 1912& ...&      	   &			        \\
NGC\,7027& 23001458& SWS01& 04-Jul-1996& 3453& ...&      	   &			        \\
NGC\,7027& 55800537& SWS01& 27-May-1997& 6537& ...&      	   &			        \\
NGC\,7027& 71300611& LWS01& 28-Oct-1997& 2156& N&        	   &			        \\
NGC\,7662& 43700427& SWS01& 26-Jan-1997& 1912& ...&      26 &	    Zhang \& Kwok (1993)  	 \\
NGC\,7662& 56801325& LWS01& 06-Jun-1997& 1318& N&        	   &			        \\
Sw\,St\,1& 47101511& SWS01& 01-Mar-1997& 1140& ...&         1.5&       Schwarz et al. (1992)	     \\
Vy\,2-2& 32002528& SWS01& 02-Oct-1996& 1140& ...&        0.6&	    Zhang \& Kwok (1993) 	 \\
Vy\,2-2& 54700310& LWS01& 16-May-1997& 1318& N&          	   &			        \\
\hline
\end{tabular}
\end{center}
\end{table*}

\twocolumn
\noindent
subsequently injected into the ISM.  Models for
soot formation in flames have been proposed, based on neutral radicals, ions,
PAHs, polyacetylenic chains, or fullerenes as intermediaries (Hucknall 1985;
Barnard \& Bradley 1985; Curl \& Smalley 1988; Ugarte 1992,1995; Hecht 1986; 
Iglesias-Groth 2004).  Each route could probably
lead to C-soot, depending on specific physical conditions.  Our goal
is to explore these routes, using C-rich planetaries as laboratories to seek
the precursors to C-dust grains.  Strong nebular UV fields excite the PAHs and
the resulting emission serves as a direct probe of the molecular gas content.

\section{The sample of Planetary Nebulae}
There are two ways to investigate the dependence of PAHs on gas-phase C/O in PNe: using 
a single well-studied band from the ground; or measuring the strongest bands from 
airborne and space-borne observatories.  An example of the former approach is the 
survey of the 3.3-$\mu$m band by Roche et al. (1996), who used common instrumentation
for their entire sample of PNe.  Previous space-based and airborne work on the 
relationships between the PAH bands and nebular C/O has depended upon the use of 
measurements made with a variety of instruments, with different apertures, and at 
different spectral resolutions.  This approach encounters the problem of poor
sensitivity to either the 11.3-$\mu$m band from the Kuiper Airborne Observatory 
(KAO) or to the 3.3-$\mu$m band from the ISO SWS.
It is clearly important to cover a wide range of IR wavelengths with 
common instrumentation for this work, so that any trends found are not compromised.  
ISO afforded a unique opportunity to secure the requisite measurements.

However, such observations were clearly predicated primarily on the IR brightness of 
the nebulae, without specific regard to their C/O ratios.  Therefore, the PNe we 
have selected have well-determined C/O values, with broad agreement among several
authors and different analytical methodologies (e.g. Kingsburgh \& Barlow
1994; Rola \& Stasinska 1994; Zuckerman \& Aller 1986).  Further, we have isolated 
precisely those nebulae whose C/O values bear most directly on the investigation of 
the trends with PAH band strength found in the old airborne data.

Table~1 summarizes the set of PNe assembled for this study; objects are listed alphabetically.  
For each PN spectrum we tabulate the name, TDT (ISO's observation identification number), the
astronomical observing template
(AOT), date, integration time, whether an LWS01 spectrum was obtained in an ``off" position, the 
angular diameter in arcsec, and a reference for the adopted diameter. 
We initially selected all PNe for which AOTs 
SWS01 and LWS01 (low-resolution spectra) were obtained by ISO.  The prerequisite 
for PN selection was the existence of a published value of C/O (eliminating 
such objects as M\,2-43 with its bright PAH emission but no C/O abundance ratio).  
The SWS aperture varies with 
wavelength.  From 2-12~$\mu$m, the size is between 14$^{\prime\prime}$ and 20$^{\prime\prime}$ 
across the several bands; from 12-29~$\mu$m, between 14$^{\prime\prime}$ and 27$^{\prime\prime}$; 
and from 29-45~$\mu$m, between 20$^{\prime\prime}$ and 33$^{\prime\prime}$.
The sample includes three large PNe that extend substantially outside the SWS apertures.  One of
these is even larger than the LWS apertures (NGC\,5189).  
For several nebulae of interest (i.e. with 
sizeable C/O) no LWS01 spectra were taken by ISO because of far-infrared (FIR) faintness (e.g. 
M\,4-18) yet the PAHs were well-detected by the SWS.  Therefore, we extended the 
sample to PNe for which SWS01 and $IRAS$ 60/100-$\mu$m photometry were available, but
which lacked any usable LWS01 spectra (i.e. absent, or too noisy).

The restriction to PNe for which SWS01 spectra are available through the ISO
Data Archive leads to a bimodal sample of nebulae.  Both PNe with bright
emission lines and very weak IR continua, and PNe of types known to exhibit 
PAH emission bands with strong IR continua (e.g. the [WCL] sequence whose 
central stars show Wolf-Rayet-like emission spectra) were almost equally 
frequently targeted by the observing community.  The integration times used for the 
sample varied by a factor of six, depending on the goals of the original
observers.  Consequently, we can offer no statement 
as to the completeness of our sample based upon any {\it physical} characteristic.

For two PNe multiple data sets exist because those objects had been
designated as LWS wavelength calibrators (NGC\,7027 (26 spectra) and NGC\,6543
(94)).  For these objects, a single representative LWS01 spectrum was chosen.
NGC\,6543 was also observed on five occasions in the SWS01 AOT but we coadded
all five spectra using inverse-variance weighting to enhance the signal-to-noise ratio.

\section{The available spectra}
The two relevant ISO spectroscopic instruments are the SWS (de Graauw et al. 1996)
and LWS (Clegg et al. 1996).  In each case we sought
low-resolution data from the AOTs ``SWS01" (covering the range from 2.38 to 
45.2~$\mu$m), and ``LWS01" (43-197~$\mu$m).

The SWS01 data include all the known mid-infrared PAH emission bands and emission plateaus
(3.2-3.6, 6-9, 10.5-15, and 16-21~$\mu$m) between 3 and 21~$\mu$m.  The SWS apertures 
were generally well-matched to the optical and radio diameters of the selected PNe. 
Typical achieved resolving powers were several hundred for AOTs taken at
the fastest speed (i.e. shortest observing time $\sim$1100s), although the 
data archive includes higher resolution spectra for some of our chosen PNe 
that were observed at the slowest speeds (e.g. resolving power 
$\sim$1500 for $\sim$6500s observing time).

LWS01 AOTs serve to assess the luminosity of the commonly found thermal 
emission from cool dust grains in PNe.  The LWS01 spectra have a resolution 
of 0.3~$\mu$m from 43--93~$\mu$m, and 0.6~$\mu$m from 84--197~$\mu$m. 
Observing times with the LWS ranged from 640 to 3400s, through an aperture with
an effective diameter of 66--86$^{\prime\prime}$, depending on wavelength.

The Spitzer Space Telescope (SST) and its infrared spectrometer (IRS) are able
to observe PNe.  However, the great sensitivity of this observatory
and its instruments implies that many of the known PNe with PAH emission are
too bright for spectroscopy with the SST.  Further, the small slits of the IRS 
cannot accommodate most of our target PNe.
Consequently, the ISO Data Archive offers an opportunity to revisit many of the 
PNe known to show PAH emission, with apertures that generally are sufficiently large to
capture essentially all the PAH emission across these objects (Smith, Aitken \&
Roche 1989) or from their surrounding photodissociation regions (PDRs) (Aitken \& Roche 1983).

We visually inspected all SWS01 spectra for PNe that survived our selection criteria, 
seeking detections of the PAH features.  Nebulae clearly showing the bands were investigated
first.  Subsequently we examined the remaining PNe with the goal of setting quantitative 
upper limits on their PAH emission.

\section{Band integrals and the join of SWS and LWS spectra}
\begin{figure}
\centering
\vspace{7cm}
\includegraphics{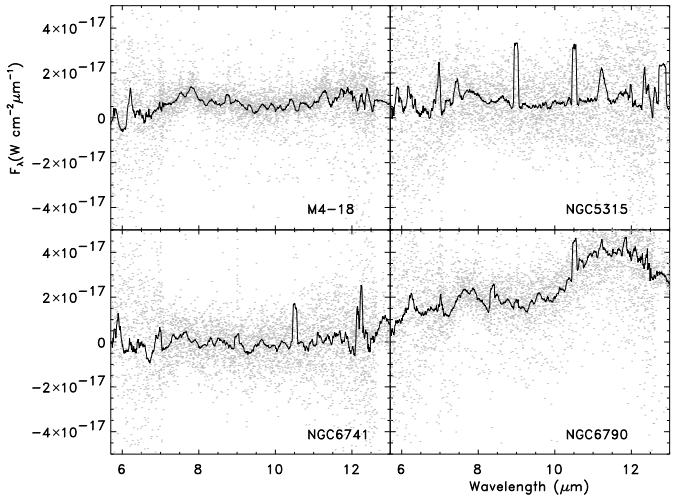}
\caption{Representative plots of the SWS region containing the chief PAH features for
PNe in our sample.  Dots show the individual observed points.
The solid curves represent the smoothed spectra.}
\label{plot1}
\end{figure}

\begin{figure}
\centering
\vspace{7cm}
\includegraphics{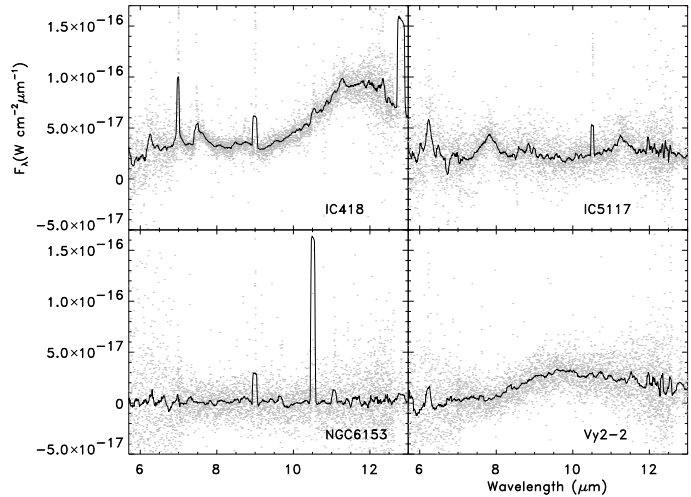}
\caption{Additional representative PNe SWS spectra.  Details as in Figure~~\ref{plot1}}. 
\label{plot2}
\end{figure}

The PAH bands are low-resolution features.
Therefore, to enhance the detectability of the bands with SWS spectra, we 
interactively used boxcar-smoothing (with widths of 50 or 75 points) 
better to define the overall continuum and PAH features in noisy data.  
Figures~\ref{plot1} and \ref{plot2} illustrate the results of this smoothing
applied to eight PNe.  The 7.7-$\mu$m and 11.3-$\mu$m PAH bands were integrated after 
interactively defining a single underlying continuum in each PN across the $5-15~\mu$m 
range.  The wavelengths selected to define these 
continua (by cubic splines) and the regions chosen for the band integrals were those
described by Cohen et al. (1986,1989) and Volk \& Cohen (1990), for which choices the 
original relationships between integrated 7.7 or 11.3-$\mu$m intensity and C/O were 
found. Care was taken to remove any influence of smoothed emission lines on these
band integrals.  For example, the 7.46-$\mu$m Pf$\alpha$ line was cut out of those nebulae
in which it was detected, by replacing it by a local smooth continuum at
the base of the line in the original, unsmoothed, SWS spectra.  Likewise, in
high-excitation PNe, we expunged the [Ne{\sc vii}] line at 7.65~$\mu$m prior
to smoothing and integrating the emission bands.  In the absence of any 
visual recognition of the PAH features, formal upper limits were set by one of
two methods.  We measured the nearest positive emission hump lying within the 
wavelength interval for definition of a PAH integration.  In addition we computed the formal 
integral over the same wavelength range of the mean spectrum plus 3$\times$ the standard 
deviation calculated over the interval.  Differences arose 
between these two methods primarily when a spectrum
had negative values or when the splined continuum locally exceeded the measured spectrum.
We adopted whichever approach yielded the more conservative upper limit.  

To evaluate the total, integrated, observed IR energy, I(IR), we assembled complete SWS+LWS 
energy distributions by splicing 
together the two data sets for each PN in the overlap region from 40-45~$\mu$m.  From 
past experience of this procedure (e.g. Cohen et al. 2002) we have determined that 
LWS01 spectra are better absolutely calibrated in this region than SWS. 
Moreover, if a PN extends slightly outside the SWS aperture, the larger LWS
aperture will more reliably assess the total nebular continuum. Therefore, 
we visually examined the overlap and rescaled each SWS spectrum upwards to achieve
the best match to the LWS when necessary.  This approach assumes that the PN emission 
lying outside the SWS Band-4 aperture is spectrally identical to that
sampled within the aperture.  

Given the complete spliced energy distribution, each spectrum was examined overall 
to eliminate any remaining specious features at the long end of the SWS band-4 range 
and/or the beginning of the LWS detector SW1 range.  The resulting cleaned and 
spliced spectra were then merged.  When a PN had multiple spectra these were combined 
using inverse-variance weighting, before splicing the resulting SWS and LWS data. 

\begin{figure}
\centering
\vspace{6.7cm}
\includegraphics{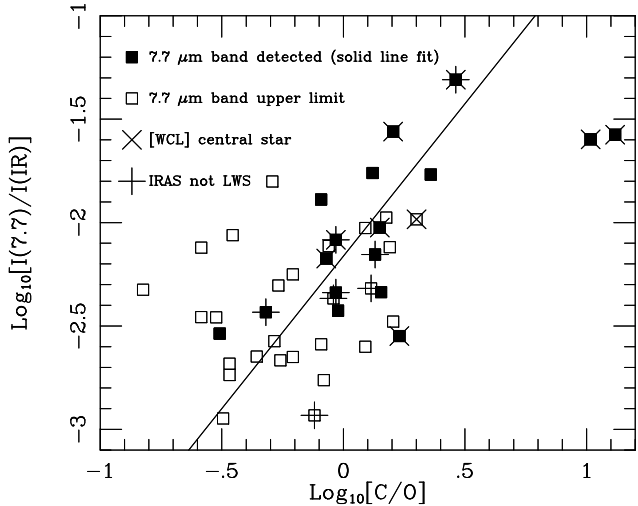}
\caption{Logarithmic plot for 43 PNe of the ratio I(7.7)/I(IR) against nebular 
gas-phase C/O abundance ratio.  Solid line is the regression for the 15 detected
PNe after excluding CPD-56$^\circ$8032 and He\,2-113.}
\label{ratio7}
\end{figure}

\begin{figure}
\centering
\vspace{6.7cm}
\includegraphics{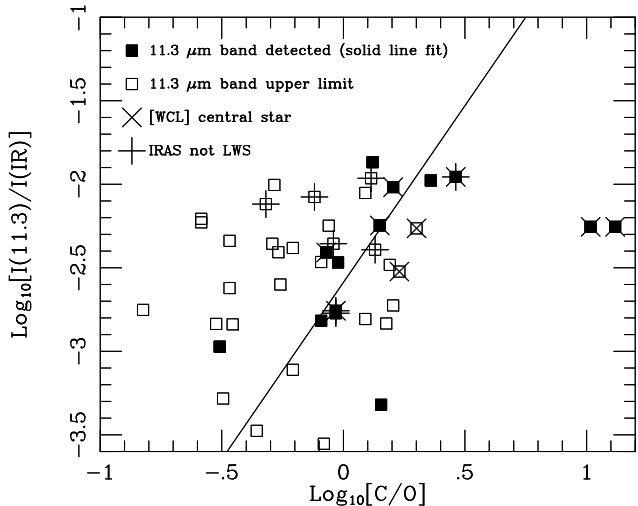}
\caption{Logarithmic plot for 43 PNe of the ratio I(11.3)/I(IR) against nebular  
gas-phase C/O abundance ratio.  Solid line is the regression for the 12 detected
PNe after excluding CPD-56$^\circ$8032 and He\,2-113.}
\label{ratio11} 
\end{figure}

For the ten PNe lacking adequate LWS01 spectroscopy, but having $IRAS$ FIR photometry, we
used 60 and 100-$\mu$m flux densities as a substitute for absent, or an alternative to
noisy, LWS01 data to estimate the FIR contribution to I(IR).  We summed the products of flux 
density and bandwidth at 60/100~$\mu$m, using the synthetic contiguous bandwidths of Emerson 
(1988) to avoid the uncertainties involved in color correction for PNe due to emission lines.
Upper limits at either 60 or 100~$\mu$m were treated as actual flux densities 
to assess the calculated FIR component of the IR luminosity.  (These ten PNe are
identified in Figures~\ref{ratio7} and \ref{ratio11} by having a large plus sign through their
plot symbols, and by asterisks in Table~2.)  
Tests were made on several PNe for which we had LWS01 data, comparing
I(IR) derived by the two methods.  These indicated that using $IRAS$ data in this fashion
produced values of I(IR) within 25~percent of the actual integrated SWS+LWS spectra.

\section{Results}
Table~2 presents the results, listing PN name, gas-phase C/O abundance ratio, I(IR) 
(in W~cm$^{-2}$), I(7.7)/I(IR), I(11.3)/I(IR), and a reference for the C/O value adopted.
Whenever possible, we have selected abundance ratios derived from collisionally 
excited forbidden lines (CELs) rather than from optical recombination lines (ORLs).  
PNe often yield substantially different abundance ratios when determined 
by the two methods (e.g. Liu et al. 2000,2001a; Tsamis et al. 2004). Liu et al. (2000) have
argued for the existence of H-deficient ionized clumps in PNe, cool enough
to suppress forbidden CELs but produce strong ORL emission.  Most of the PNe 
in Table~1 have CEL C/O ratios.  When these are unavailable, we have used the
ORL line ratio (e.g., for Vy\,2-2).  We have never mixed CEL and ORL 
abundance determinations to create a C/O value.

In the first studies of the multi-band PAH spectra of PNe from the KAO it
was recognised that the ratio of I(7.7)/I(IR) ranged over two orders of      
magnitude.  To avoid confining the distribution of the bulk of the PNe 
detected to the lower left corner of any figure against C/O by use of a linear 
plot, the logarithmic ordinate was selected.  We have followed this precedent.
CPD-56$^\circ$8032 and He\,2-113 are well-measured objects with extreme values of C/O, 
yet they both emit comparable fractions of the IR flux in the PAH bands to those emitted 
by other PNe with much more modest C/O ratios.  Consequently, we exclude them from  
our sample in terms of any regression analysis.  

Fig.~\ref{ratio7} illustrates the 7.7-$\mu$m results, distinguishing between
PNe detected in the 7.7-$\mu$m band, the subset of nebulae with [WCL] central 
stars, and the 26 objects with only upper limits.  The solid line plotted in this figure
indicates the formal regression line for all detected PNe with the exceptions of
CPD-56$^\circ$8032 and He\,2-113.  The intercept is $-2.16\pm0.06$, the slope is
1.47$\pm$0.33, and the Pearson correlation coefficient for these 15 PNe is {\it r}=0.76.
We identify this dashed regression line as the ISO version of the relationship reported
by Cohen et al. (1989: their Figure~20) based on airborne spectroscopy.  (Note
that, in their figure, the point for CPD-56$^\circ$8032 was misplotted too high 
by a factor of 10 in ordinate.  It should have appeared at -1.1.)  

The regression line for the six [WCL] objects (after excluding CPD-56$^\circ$8032 and 
He\,2-113) with the 7.7-$\mu$m band is 
insignificantly different (at the 1$\sigma$ level) from the line for the 15 PNe.
M\,4-18 has the largest value of I(7.7)/I(IR).  Ground-based
spectra (Aitken \& Roche 1982; Rinehart et al. 2002) clearly show its 11.3-$\mu$m band
although its 8.7-$\mu$m feature is convincingly shown only by the Aitken \& Roche (1982) 
spectrum from Mauna Kea.  M\,4-18's SWS spectrum is very noisy, although the PAH features 
are unquestionably detected.  However, the spectral levels of the PAH bands in this SWS
spectrum match those in the spectrum by Rinehart et al. (2002) to within 10~percent,
confirming our estimate of I(7.7)/I(IR) for this object.
In the small sample of PNe studied by Cohen et al. (1989) NGC\,6302 substantively
helped to define the trend at 7.7~$\mu$m because of its low C/O ratio.
With our enlarged set of PNe NGC\,6302 no longer controls the existence of a trend
in this diagram.  The correlation coefficient remains {\it r}=0.76 even if one were
to exclude NGC\,6302 along with CPD-56$^\circ$8032 and He\,2-113.  There is no cause
to reject this PN; we simply make the point that a single PN does not strongly influence 
the regression line in this sample at 7.7~$\mu$m.

One immediately sees that the [WCL] PNe dominate the plot by contributing almost all the 
large ratios of I(7.7)/I(IR).  Only NGC\,6369, among the [WCL] PNe, was not
detected in PAH emission.  Roche, Aitken \& Whitmore (1983) observed this PN from 
the ground at very low resolving power ($\sim$40), in a 20$^{\prime\prime}$ beam 
which accommodated about half the ionized gas distribution.  At best there are
possible suggestions of emission humps near 8.7 and 11.3-$\mu$m emission in their 
spectrum.  The object has only a single SWS01 spectrum, as opposed to several of 
the [WCL] nebulae which have higher IR surface brightness.

All the 7.7-$\mu$m non-detections have relatively short exposure (fast speed) SWS01 
observations (1062-1912\,s), suggesting that they were spectra with poor signal-to-noise.
Almost all PNe with C/O ratios $\geq$0.85 (log$_{10}~$C/O~$\geq$-0.07)
and I(7.7)/I(IR) $>$0.4~percent (log$_{10}$~C/O~$\geq$-2.45) were detected in this PAH feature.  
Seven PNe, lying in the lower-left quadrant of Fig.~\ref{ratio7}
below these limiting values, were not detected in the 7.7-$\mu$m band.  
Only NGC\,6302 is detected outside the above limits of C/O and I(7.7)/I(IR).
This detection was certainly aided by the use of the slowest speed SWS01 AOT (6528\,s)
because the PAHs are weak in this PN and I(IR) is large.   

Fig.~\ref{ratio11} similarly presents the 11.3-$\mu$m results.  A single regression 
line is again plotted for the 12 detected PNe, excluding CPD-56$^\circ$8032 and He\,2-113.
The intercept is -2.59$\pm$0.08 and slope 2.12$\pm$0.49.  The influence of
NGC\,6302 on this regression line is again minimal: {\it r} drops from 0.71 to 0.67
if one were to exclude this object from the regression analysis.  Clearly there is no
justification for doing this.  The PN furthest from the line, with the lowest I(11.3)/I)IR) 
of the sample, is IC\,418 although there is likewise no reason to exclude it.  Several
nebulae show a broad hump between 10.6 and 12.4~$\mu$m attributed to emission by SiC
grains.  In these objects the 11.3-$\mu$m PAH band is a very small emission feature
near the peak of the SiC structure, requiring careful definition of a local continuum
to extract the PAH feature.  Perhaps this accounts for its unusual weakness in
IC\,418, whose SiC emission is very bright.
Overall, I(11.3)/I(IR) rises abruptly at low C/O and maintains this rise at least as far
as C/O $\sim2$, exactly as found by Volk \& Cohen (1990: their Figure~3).  Only
the two extreme PNe are observed beyond this value of C/O.

\begin{figure}
\centering
\vspace{6.7cm}
\includegraphics{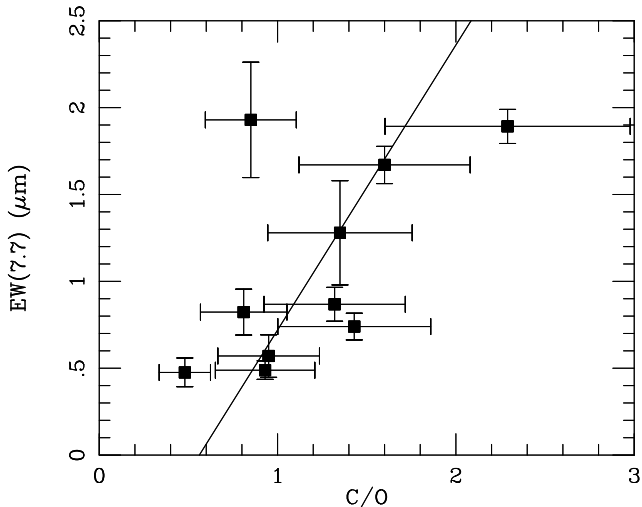}
\caption{Linear plot of the EW(7.7) against nebular gas-phase C/O abundance ratio.  
Solid line is the regression for the 10 detected PNe for which meaningful EWs can
be measured after excluding CPD-56$^\circ$8032 and He\,2-113.}
\label{ew77}
\end{figure}

From their analysis of the 3.3-$\mu$m band, Roche et al. (1996) suggested that the
equivalent width (EW) of this band argued for a cut-off below C/O~=~0.6.  For 
high quality near-infrared ground-based spectra, EWs are appropriate.  For space-based 
spectra the sample is still limited by signal-to-noise.  This can lead to poor estimates 
of the local continuum required to define an EW.  Therefore, I(7.7)/I(IR) is a far more 
robust measure of PAH strength than EW for our sample of PN spectra.  However, in the 
interest of trying to determine whether our sample supports a similar conclusion, we have 
reformulated the 7.7-$\mu$m data as EWs.  But this lack of robustness means that we must
exclude several PNe from the sample.  These are precisely those objects in which portions 
of the overall splined continua are so noisy that negative continua result at some 
wavelengths within the broad 7.7-$\mu$m feature, or else very large EWs ($>$2~$\mu$m) are 
formally derived.  We again exclude CPD-56$^\circ$8032 and He\,2-113 but must also
reject Sw\,St\,1, NGC\,40, NGC\,6302, Cn\,1-5, and M\,4-18.  The final sample includes 10
PNe.  We have carried out a careful estimation of the errors in the derived EWs by adding
in quadrature the uncertainties in each integrated 7.7-$\mu$m intensity with those associated
with the definition of the underlying continuum.  Fractional uncertainties of 30\%
have been assigned to the observed C/O values.
Fig.~\ref{ew77} presents the EWs for these PNe linearly against C/O.  The solid
line is the best fitting relationship.  The formal detectable onset of 7.7-$\mu$m emission
is for C/O~=~0.56$^{0.21}_{0.41}$, which includes the value of 0.6 found by Roche et al. (1996), 
at which C/O value they estimate that 3.3-$\mu$m PAH emission first appears.

Comparing the behaviour of CPD-56$^\circ$8032 and He\,2-113 with the regressions in
Figures~\ref{ratio7} and \ref{ratio11} suggests that perhaps these two objects are so
extreme because some process ``saturates" so that further increase of C/O no longer 
yields more intense PAH emission relative to the overall dust continuum emission.
For both these PNe, essentially all of the stellar 
ultraviolet (UV) photons are being absorbed and reradiated in the IR by the PAHs
and dust particles (Aitken et al. 1980), which is not the case for most of
the other objects in our sample.  Around these two objects, there could
therefore be a significant stratification with radius of both the spectral
quality of the ambient UV radiation field, with harder photons being
absorbed closer in, and of the characteristic grain temperature, so
that strong PAH emission closer in may be offset by weak PAH and
strong far-IR grain emission further out. Therefore, the best guides
to any underlying correlations between band strengths and C/O ratios
are those provided by our PN samples that omit CPD-56$^\circ$8032 and He\,2-113.

The 6.2-$\mu$m PAH band is seen in 14 PNe and their relative strengths, I(6.2)/I(IR), are given
in Table~3.  In most of these nebulae this band is measured rather poorly because high
noise in the SWS spectra below 5-6~$\mu$m adversely affects the definition of the underlying
continua that we subtract.  There is no meaningful correlation between I(6.2)/I(IR) and C/O.

\begin{table*}
\begin{center}
\caption{Spectrally integrated observed IR fluxes, I(IR) (in W~cm$^{-2}$), and the ratios I(7.7)/I(IR) and I(11.3)/I(IR), 
for 43 PNe. Names with asterisks denote PNe with I(IR) based on the SWS data and $IRAS$ FIR photometry.}
\begin{tabular}{crrrrl}
\hline
PN&     C/O&    I(IR)& I(7.7)/I(IR)& I(11.3)/I(IR)& C/O Reference\\
\hline        
BD+30$^\circ$3639&    1.60& 4.90E-15& 0.0275&    0.0096&   Pwa, Pottasch \& Mo (1986)\\
CPD-56$^\circ$8032&   13.1& 7.77E-15& 0.0266&    0.0056&   De Marco, Barlow \& Storey (1997)\\
Cn\,1-5&              1.7 & 4.77E-15& 0.0028&    $<$0.0030&  Walton, Barlow \& Clegg (1993)\\
Hb\,12$^*$&           0.52& 1.05E-15& $<$0.0027& $<$0.0099&   Hyung \& Aller (1990)\\ 
He\,2-113&            10.4& 6.00E-15& 0.0253&    0.0056&   De Marco, Barlow \& Storey (1997)\\
He\,2-131&            0.30& 1.60E-15& $<$0.0035& $<$0.0015&  Liu et al. (2001a)\\
Hu\,1-2$^*$&          1.23& 8.60E-17& $<$0.0094& $<$0.0089&   Liu et al. (2004)\\
Hu\,2-1$^*$&          0.48& 1.80E-16& 0.0037&    $<$0.0076&  Wesson et al. (2005)\\
IC\,3568 &            0.54& 1.49E-16& $<$0.0050& $<$0.0039&   Liu et al. (2004)\\
IC\,418  &            1.43& 2.38E-15& 0.0046&    0.0005&   Henry, Kwitter \& Bates (2000)\\ 
IC\,4406 &            0.62& 2.37E-16& $<$0.0022& $<$0.0042&  Tsamis et al. (2003)\\
IC\,4997 &            0.34& 2.31E-16& $<$0.0021& $<$0.0046&  Hyung et al. (1994)\\ 
IC\,5117 &            1.32& 8.29E-16& 0.0174&    0.0136&   Hyung et al. (2001)\\ 
M\,1-42  &            0.15& 1.29E-16& $<$0.0047& $<$0.0018&   Liu et al .(2001b)\\ 
M\,2-36  &            0.76& 9.15E-17& $<$0.0012& $<$0.0084&   Tsamis et al. (2003)\\
M\,4-18$^*$&          2.9 & 1.80E-16& 0.0491&    0.0111&   De Marco \& Barlow (2001)\\ 
Mz\,3    &            0.83& 7.42E-15& $<$0.0017& $<$0.0003&  Zhang \& Liu (2002)\\ 
NGC\,2346&            0.35& 1.15E-16& $<$0.0087& $<$0.0015&   Rola \& Stasinska (1994); Peimbert \& Torres-Peimbert (1987)\\
NGC\,2440$^*$&        0.91& 4.49E-16& $<$0.0043& $<$0.0044&  Hyung \& Aller (1998)\\ 
NGC\,3918&            1.60& 9.59E-16& $<$0.0033& $<$0.0019&  Ercolano et al. (2003)\\
NGC\,40&              1.41& 1.38E-15& 0.0095&    0.0057&  Liu et al. (2004)\\
NGC\,5189&            0.34& 1.71E-16& $<$0.0018& $<$0.0024&  Liu et al. (2001a)\\
NGC\,5315&            0.85& 1.12E-15& 0.0067&    0.0039&   Pottasch et al. (2002a)\\ 
NGC\,6153&            0.55& 1.20E-15& $<$0.0022& $<$0.0025&   Liu et al. (2000)\\ 
NGC\,6210&            0.26& 4.21E-16& $<$0.0035& $<$0.0059&   Liu et al. (2004)\\
NGC\,6302&            0.31& 8.08E-15& 0.0029&    0.0011&  Tsamis et al. (2003)\\ 
NGC\,6369&            2.00& 1.54E-15& $<$0.0104& $<$0.0055&  Zuckerman \& Aller (1986);Aller \& Keyes (1987)\\
NGC\,6445&            0.81& 3.38E-16& $<$0.0026& $<$0.0034&   Pottasch et al. (2002b)\\ 
NGC\,6537&            0.95& 1.50E-15& 0.0038&    0.0034&  Liu et al. (2001)\\
NGC\,6543&            0.44& 2.19E-15& $<$0.0023& $<$0.0003&  Wesson \& Liu (2004)\\
NGC\,6572&            1.55& 1.55E-15& $<$0.0076& $<$0.0033&   Liu et al. (2004)\\
NGC\,6720&            0.62& 2.64E-16& $<$0.0056& $<$0.0008&  Liu et al. (2004)\\
NGC\,6741$^*$&        1.35& 2.78E-16& 0.0070&    $<$0.0041&  Liu et al. (2004)\\
NGC\,6790&            0.81& 6.66E-16& 0.0129&    0.0015&  Liu et al. (2004)\\
NGC\,6826&            0.87& 1.05E-15& $<$0.0077& $<$0.0057&   Liu et al. (2004)\\
NGC\,6884$^*$&        0.93& 3.03E-16& 0.0046&    0.0017& Liu et al. (2004)\\
NGC\,6886$^*$&        1.3&  2.50E-16& $<$0.0048& $<$0.0109&   Hyung et al. (1995)\\
NGC\,6891&            0.51& 1.62E-16& $<$0.0158& $<$0.0044&   Wesson et al. (2005)\\ 
NGC\,7009&            0.32& 1.13E-15& $<$0.0011& $<$0.0005&  Liu et al. (2004); Liu et al. (1995)\\
NGC\,7027&            2.29& 2.05E-15& 0.0171&    0.0106&   Middlemass (1990)\\
NGC\,7662&            1.23& 5.60E-16& $<$0.0025& $<$0.0016&  Liu et al. (2004)\\
Sw\,St\,1$^*$&        0.93& 1.47E-15& 0.0083&    0.0018&  De Marco et al. (2001)\\ 
Vy\,2-2$^*$&          0.26& 8.46E-16& $<$0.0076& $<$0.0062& Wesson et al. (2004)\\
\hline
\end{tabular}
\end{center}
\end{table*}

\begin{table*}
\begin{center}
\caption{Detected PAH 6.2-$\mu$m ratios, I(6.2)/I(IR), for PNe}
\begin{tabular}{cr}
\hline
PN&                  I(6.2)/I(IR)\\
\hline 
BD+30$^\circ$3639&   0.0073\\
Cn\,1-5&             0.0041\\
CPD-56$^\circ$8032&  0.0105\\
He\,2-113&           0.0116\\
IC\,418  &           0.0021\\
IC\,5117 &           0.0140\\
M\,4-18  &           0.0051\\
NGC\,5315&           0.0037\\
NGC\,6302&           0.0008\\
NGC\,6790&           0.0089\\
NGC\,6826&           0.0024\\
NGC\,6886&           0.0079\\
NGC\,7027&           0.0039\\
Sw\,St\,1&           0.0050\\
\hline
\end{tabular}
\end{center}
\end{table*}

\section{Conclusions}
PAHs were not detected in two of the three largest nebulae in our sample, either because the
circumnebular PDRs that ought to contain PAHs are not sampled by the modest sizes of the SWS 
apertures, or else because these lower density nebulae are optically thin to ionizing 
radiation and have no PDRs. This could explain the
absence of detectable 7.7-$\mu$m emission bands in NGC\,5189 and NGC\,6720.
PAHs were detected by the SWS in NGC\,6537, despite its diameter of 70$^{\prime\prime}$.  However,
this object is bipolar and has a similarly high excitation class and morphology to NGC\,6302, in 
which PAHs are also detected.  This morphology implies a central concentration of material that
is lacking in large classical PNe like NGC\,6720.  Indeed, the strong 3.3 and 3.4-$\mu$m bands 
seen in a 3$^{\prime\prime}$ slit spectrum of NGC\,6537 by Roche et al. (1996) offer confirmation 
that the PAH emission in at least this nebula arises in the compact core.

Using a sample of PNe more than twice as large as that previously available from airborne 
spectroscopy, we have vindicated the existence of a relationship between I(7.7)/I(IR) and 
gas phase C/O.  The fraction of total IR luminosity emitted by the 7.7-$\mu$m band is
observed to be roughly linearly proportional to C/O for abundance ratios up to $\sim$3.
From a sample of PNe about the same size as that used in the LRS study by Volk \& Cohen 
(1990) but based on higher signal-to-noise SWS spectra, we have confirmed that I(11.3)/I(IR) 
rises rapidly with C/O to at least C/O $\sim$2, growing roughly as the square of the 
gas-phase abundance ratio.  For nebulae with C/O beyond a value of 2-3 this fractional growth 
of the PAH emission bands apparently ceases.  The difference between the logarithmic 
intercepts of these two relationships (-2.59 and -2.16) indicates that, for PNe with 
C/O$\approx$1, the intensity of the 11.3-$\mu$m band is 37~percent of the 7.7-$\mu$m band's 
intensity.  Note that the use of logarithmic plots greatly reduces the dependence of this
ratio of band strengths on abundance ratio because the centroid of our sample is C/O$\approx$1.
This is in excellent agreement with the average of 36~percent measured by Cohen 
et al. (1986, their Table~5) for the generic spectrum of astrophysical PAHs.  
The 1986 sample included only 7 PNe for which band ratios relative to I(7.7) were definable
and included H{\sc ii} regions and reflection nebulae too.
We have, therefore, redetermined I(6.2)/I(7.7) and I(11.3)/I(7.7) purely from our 
new sample of PNe, and find a mean ratio and standard error of the mean of 
0.62$\pm$0.13 (13 PNe) and 0.40$\pm$0.06 (14 PNe), respectively.

These empirical correlations between PAH band strength and nebular C/O abundance ratio 
could yield insights into the carbon dust condensation process because PAHs have been 
suggested to be the likely molecular precursors of C grains (Crawford, Tielens 
\& Allamandola 1985; Allamandola, Tielens \& Barker 1989).  
I(7.7)/I(IR) essentially measures the far-ultraviolet absorption cross
sections of PAHs with respect to that of the larger C-grains, believed to
dominate the continuum emission in planetaries. One obvious interpretation
of the observed correlations in Figs.~\ref{ratio7} and \ref{ratio11} is that 
{\it n}(PAH)/{\it n}(carbon grains) increases with increasing C/O ratio. If, on the other
hand, {\it n}(PAH)/{\it n}(carbon grains) is largely invariant, one could
interpret the correlations of PAH band strengths with C/O as due to growth of the grain 
component (higher C/O leading to bigger particles with smaller
UV cross sections per unit volume, so that PAHs absorb more UV energy
relative to the grains). 
 
\section{Acknowledgements}
We thank the anonymous referee for comments that have helped improve the paper.
MC thanks NASA for supporting the early portion of this work through grant
NAG5-4884, and later through contract JPL961501, with Berkeley.
It is a pleasure to thank the Physics \& Astronomy Dept. at UCL for hosting a number 
of short-term visits that have enabled the completion of this effort.

\end{document}